# Observation of coexisting weak localization and superconducting fluctuations in strained $Sn_{1-x}In_xTe$ thin films.

*Jiashu Wang,[1]\* William Powers,[1] Zhan Zhang[2], Michael Smith,[1] Bradlee J. McIntosh,[1] Seul-Ki Bac,[1] Logan Riney,[1] Maksym Zhukovskyi,[3] Tatyana Orlova,[3] Leonid P. Rokhinson,[4,5,6] Yi-Ting Hsu,[1] Xinyu Liu,[1] Badih A. Assaf[1]\**

1. Department of Physics, University of Notre Dame, Notre Dame, IN, 46556

2. X-ray Science Division, Advanced Photon Source, Argonne National Lab, Lemont, IL

3. Notre Dame Integrated Imaging Facility, University of Notre Dame, Notre Dame, IN, 46556

4. Department of Physics and Astronomy, Purdue University, West Lafayette, IN 47907

5. Birck Nanotechnology Center, Purdue University, West Lafayette, IN 47907

6. Department of Electrical and Computer Engineering, Purdue University, West Lafayette, IN 47907

ABSTRACT. Topological superconductors have attracted tremendous excitement as they are predicted to host Majorana zero modes that can be utilized for topological quantum computing. Candidate topological superconductor $Sn_{1-x}In_xTe$ thin films (0<x<0.3) grown by molecular beam epitaxy and strained in the (111) plane are shown to host three coexisting quantum effects: localization, antilocalization and superconducting fluctuations above the critical temperature $T_c$. An analysis of the normal state magnetoresistance reveals these effects. Weak localization is consistently observed in superconducting samples, indicating that superconductivity originates dominantly from trivial valence band states that may be strongly spin-orbit split. A large enhancement of the conductivity is observed above $T_c$, indicating that quantum coherent quasiparticle effects coexist with superconducting fluctuations. Our results motivate a re-examination of the debated pairing symmetry of this material when subjected to quantum confinement and lattice strain.

MAIN TEXT

Non-trivial topology can arise in electronic band structures and lead to unique properties in various materials that include insulators, semimetals, magnets and superconductors.[1,2,3,4] In particular, superconductors that are topological are proposed to have a gapped bulk, accompanied by gapless boundary states dubbed Majorana boundary modes. These exotic excitations can have possible applications in quantum computing, [5] which greatly motivates the current interest in topological superconductors (TSCs). TSCs can arise in materials that are topologically non-trivial and superconducting.[6] Materials proven to be intrinsic TSCs are rare but candidates include $Cu_xBi_2Se_3$,[7,8] $UTe_2$,[9] $UPt_3$ [10] and $Sn_{1-x}In_xTe$. Alternatively, one can achieve TSCs employing the proximity effect;[11] namely by building heterostructures of topological insulators (TIs) and conventional superconductors, one introduces superconductivity into TIs[12,13,14]

SnTe is a topological crystalline insulator (TCI).[15,16] It hosts four Dirac cones on the (111) surface protected by crystal mirror symmetry.[17,18] $Sn_{1-x}In_xTe$(SIT) can become superconducting while maintaining its topological surface states when alloyed with In.[19,20,21,22,23] However, the non-trivial character of superconductivity in $Sn_{1-x}In_xTe$ remains debated, and several experiments, including recent scanning electron spectroscopy measurements could not reveal the presence of odd pairing or Majorana modes.[24,25,26,27,28] Regardless of whether or not it is intrinsically topological, $Sn_{1-x}In_xTe$ is one of the few superconductors that are epitaxially compatible with TCIs (Pb,Sn)Te and can be used in proximity effect heterostructures based on $Sn_{1-x}In_xTe$/(Pb,Sn)Te to achieve the TSC.

Most previous research on $Sn_{1-x}In_xTe$ was focused on single crystals. The growth and analysis of $Sn_{1-x}In_xTe$ thin films are necessary to search for boundary Majorana modes using device-based schemes. Thin films also allow to manipulate the properties of the material using tuning knobs beyond those possible for single crystals such as confinement, strain and interfacial effects. In thin films, quantum coherent corrections to the conductivity that are useful to quantity spin-orbit splitting effects are enhanced. Moreover, in the case of superconductors, confinement enhances quantum coherent corrections due to superconducting fluctuations above the critical temperature $T_c$, which could provide a novel route to put to test the order parameter symmetry. The MBE of $Sn_{1-x}In_xTe$[29] was recently developed on InP substrates but measurements involving the effects of confinement on the superconducting and normal state properties of $Sn_{1-x}In_xTe$ are yet to be carried out.

Here, we report structural and electrical measurements on strained thin films of $Sn_{1-x}In_xTe$(0.04<x<0.3) grown by MBE on $BaF_2$ (111) that reveal remarkable differences compared to single crystals. The films host rhombohedral lattice strain along the (111) plans that alters the crystal symmetry. They are all superconducting for *x>0.04* but also exhibit strongly enhanced quantum coherent corrections to the conductivity that unravel important properties of the normal state. In the normal state, weak localization (WL), weak antilocalization (WAL) and superconducting fluctuations coexist. The WL only occurs in superconducting samples, indicating that electrons from trivial bulk states may play a dominant role in the superconducting state. However, the observation of antilocalization at low magnetic field indicates the presence of bulk spin-orbit splitting possibly resulting from a Rashba splitting whose origin is debated. Close to but above $T_c$, a significant enhancement of the conductivity and enhanced strong negative magnetoconductance indicate the presence of superconducting fluctuations of Maki-Thompson type. The absence of a signature of fluctuations in the Meissner effect corroborates this interpretation. Our realization of $Sn_{1-x}In_xTe$ thin films and the observation of the WAL and fluctuations shed a new light on the properties of this material, likely to change under the impact of quantum confinement and lattice strain. Interestingly, strain in the (111) plane breaks cubic symmetry and introduces a rhombohedral lattice deformation whose impact on the superconducting properties and pairing symmetry has not yet been studied in $Sn_{1-x}In_xTe$.

Thin film of $Sn_{1-x}In_xTe$ (0.04<x<0.3) are grown by molecular-beam epitaxy (MBE) on $BaF_2$(111) substrate. We initially deposit a buffer layer of SnTe (<40nm) to start the nucleation of the layer. We then co-deposit Sn, In and Te from elemental cells. Throughout the growth, we vary the cell temperature as well as the relative growth time of SnTe and $Sn_{1-x}In_xTe$ layers to control composition (see supplementary information). Layers of $Sn_{1-x}In_xTe$ of various concentration varying from 0 to 0.3 are obtained. The In atoms are found to diffuse into the SnTe buffer almost regardless of composition and buffer layer thickness.

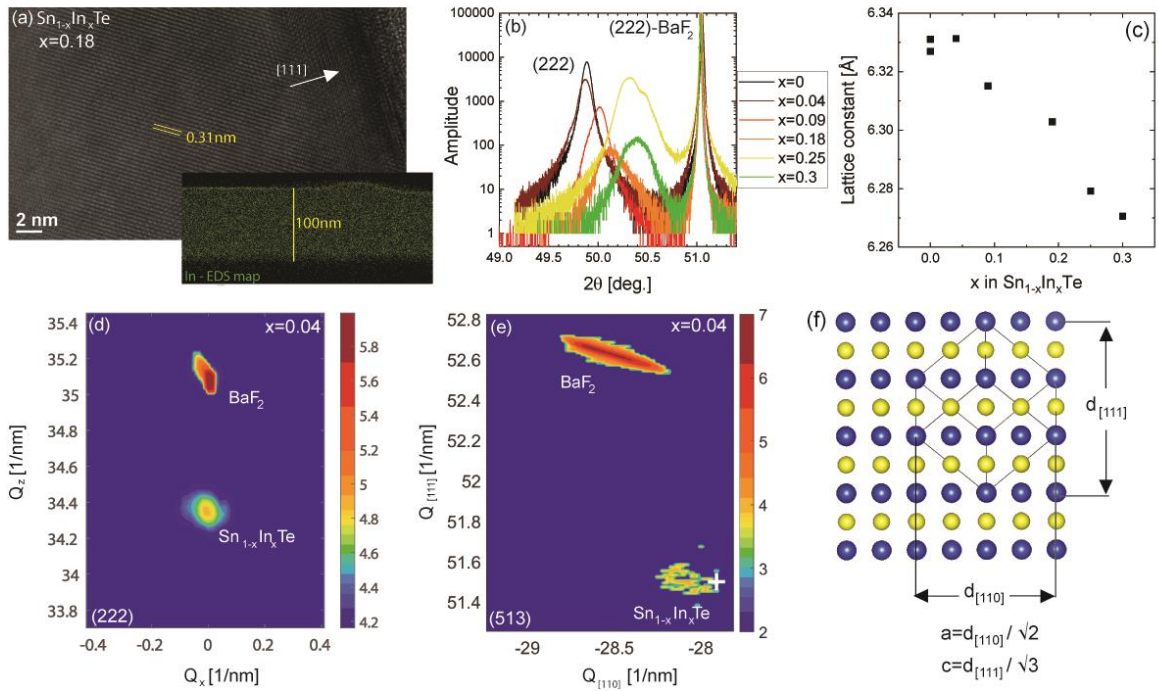

Figure 1. (a) Transmission electron microscope image of a $Sn_{1-x}In_xTe$ (x=0.25) thin film. Inset shows an elemental mapping of the energy dispersive X-ray emission line, showing a homogeneous distribution within the instrument resolution. (b) X-ray diffraction scan about the (222) Bragg peak of films having 0<x<0.3. (c) Lattice constant obtained in the (111) direction versus x. (d) RSM around (222) substrate and layer peaks. (e) RSM about the (513) peaks. The white cross corresponds to the peak expected from a relaxed cubic structure with c=a. (f) Schematic of a (111) oriented film showing a and c lattice constants obtained from the [110] and [111] lattice spacing, respectively.

Transmission electron microscopy (TEM) measurements performed at room temperature on a sample having x=0.18 confirm this. A TEM image of the layer is shown in Fig. 1(a), along with an energy dispersive X-ray (EDX) map of the In line in the inset. The image yields well resolved atomic layers confirming the good crystalline quality of the films. The EDX measurements show a uniform In distribution within the resolution of the instrument (~±2%). The diffusion of In during the growth thus leads to a homogeneous $Sn_{1-x}In_xTe$ layer with a thickness of 100nm.

X-ray diffraction scans of the (222) Bragg peaks of various $Sn_{1-x}In_xTe$ films are shown in Fig. 1(b). The peak position shifts to higher angle with increasing In indicating decrease of the lattice constant as In is alloyed into the structure. This trend is plotted in Fig. 1(c) where between *x=0.04* and *x=0.3*, a clear decrease of the lattice constant is obtained versus temperature. This agrees with previous reports on single crystals and thin films of $Sn_{1-x}In_xTe$.[29,30]

The slight lattice mismatch of 1% to 2.2% between $Sn_{1-x}In_xTe$ (~6.33 – 6.26Å) and $BaF_2$ (6.196Å) warrants a thorough investigation of the in-plane lattice parameter. Thus, high-resolution X-ray diffraction is performed around the (222) and (513) Bragg peaks. The resulting reciprocal maps (RSMs) for x=0.04 are shown in figures 1(c) and (d), respectively. The analysis of the RSMs yields the lattice constant c and a, respectively in the directions parallel and perpendicular to the [111] growth direction. Here, we define $c = d_{[111]}/\sqrt{3}$ and $a = d_{[110]}/\sqrt{2}$ where $d_{[hkl]}$ represents the lattice spacing in the [hkl] direction. From Fig. 1(c), we use the (222) Bragg peak to obtain c=6.331Å (±<0.001). From Fig. 1(d), the 513 peak allows us to find a=6.290±0.003Å (the uncertainty spans the width of the peak and is thus an overestimation). The layer is thus compressively strained in the (111) plane and tensile-strained out-of-plane along the growth axis, thus breaking cubic crystalline symmetry (Fig. 1(f)). Similar data for x=0.09 is shown in the supplement. In Fig. 1(e), a cross locates the peak expected from an unstrained layer (i.e. a=c) on the RSM. The experimental Bragg peak is broad but evidently located left of the cross. Thus, even at 100nm, the layer remains under stress and c>a.

Electrical transport measurement are performed from 1.5K to 15K using the standard 5-probe method. Gold wires are attached to the film using conducting silver paste. All In-containing samples exhibit a superconducting regime at low temperature. As we increase the In concentration, the critical temperature $T_c$ gradually rises from 1.5K to 3.5K, as shown in Fig. 2(a), in agreement with previous work on single crystals and thin films.[29,31,32,33,21] Figures 2(b,c) plot the sample resistance versus magnetic field applied out-of-plane and in-plane respectively. The plots show several temperatures below $T_c$. The upper critical field $H_{c2}$ defined as the mid-point between the normal state resistance and 0 is extracted and plotted versus temperature in Fig. 2(d). The plots show the anisotropy between out-of-plane and in-plane directions and a quasi-linear relationship of $H_{c2}$ with respect to temperature, for both directions. This indicates that the superconducting state is quasi-3D. The two curves can be well fit by the 3D Ginzburg-Landau theory:

$$\mu_0 H_{c2}^i = \mu_0 H_{c2}^i(0)\left[1 - \frac{T}{T_c}\right]$$

Where $H_{c2}^i(0) = H_{c2}^{\perp,\|}(0)$ are the critical field at T=0K, found to be $H_{c2}^\perp(0) = 1.92T$ and $H_{c2}^\|(0) = 3.27T$. This is below the Pauli limit $B_P \approx 1.85(TK^{-1}) \cdot T_C \approx 6.5T$. Using the relation $\mu_0 H_{c2}^{\perp}(0) = \frac{\Phi_0}{2\pi\xi_{ab}^2}$ and $\mu_0 H_{c2}^{\parallel}(0) = \frac{\Phi_0}{2\pi\xi_{ab}\xi_c}$, where $\Phi_0 = \frac{h}{2e}$ is the flux quantum, we can estimate the superconducting coherence length $\xi_{ab}$(in plane) and $\xi_c$(out of plane). The extracted values are 12.9nm and 7.6nm separately. Their value is smaller than the thickness of the film (100nm), meaning the order parameter is well confined in the film, leading to quasi-3D superconductivity.

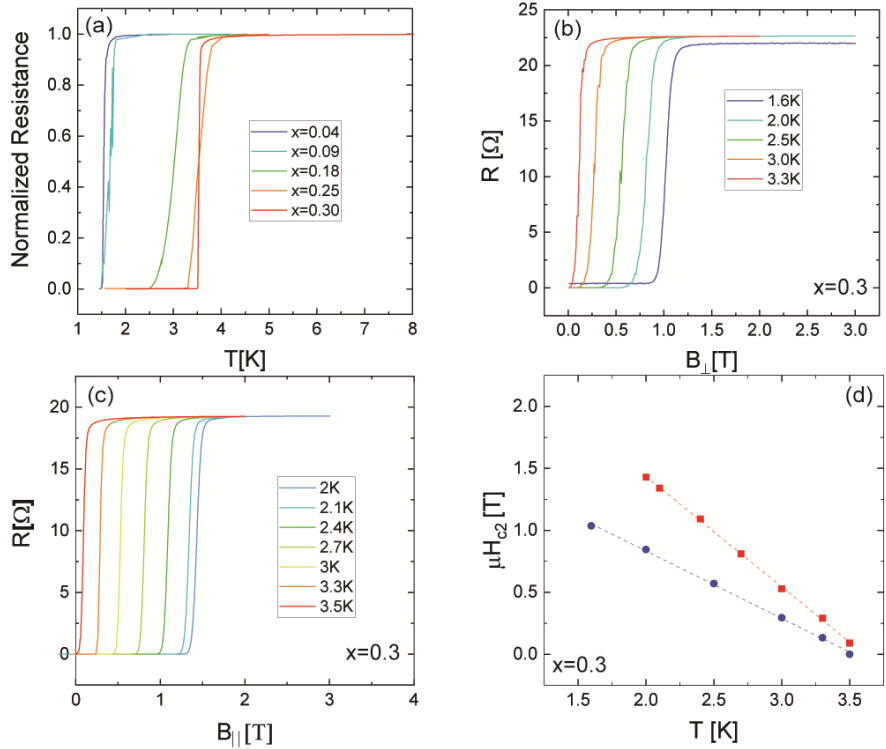

Figure 2. (a) Dependence of $T_c$ with In concentration. (b) Magnetic field dependence of resistivity in the superconducting regime for $Sn_{0.7}In_{0.3}Te$ at different temperature for a field applied perpendicular to the sample plane. (c) Same as (b) but for field applied in the plane perpendicular to the current. (d) Temperature dependence of critical field applied in-plane and out-of-plane. Points are taken at half of the normal state resistance. Dashed lines are fitting curves using Ginzburg-Landau theory.

Fig. 3 summarizes the transport measurements carried out in the normal state. The Hall effect, shown in figure 3(a), reveals a decreasing Hall constant with increasing In, suggesting that In is an acceptor. However, the carrier density calculated based on this exceeds $10^{23}$ cm$^{-3}$ for x≥0.18 which is unrealistic as it exceeds the atomic density. This corresponds to a Hall coefficient ~0.0001cm$^3$/C. We suspect that the small Hall coefficient can be caused by the coexistence of holes and electrons that both have a low mobility as previously discussed in single crystals.[34] In this situation, the Hall resistivity in the Drude model follows:

$$\rho_{xy} \approx \frac{(n_h - n_e)}{e(n_h + n_e)^2} B$$

Where $n_h, n_e$ is the carrier density of electrons and holes and B is the magnetic field. Thus, the effective carrier density calculated from Hall slope can be unreasonably high if $n_h - n_e$ is small. As we increase In, $n_e$ increases, thus gradually lower the Hall coefficient. From this, we conclude that the In acts as an electron donor, possibly generating an impurity band near the Fermi level. Several recent works also report similar observation[23,34]. This impurity band is believed to play an essential role in the superconductivity of this material, particularly if a mixed valence state can be revealed for In in SnTe.[19]

In the normal state, for all superconducting samples, a cusp at lower perpendicular field followed by a negative magnetoresistance at intermediate field are observed as shown in figure 3(b,c). Such a behavior can generally be attributed to the coexistence of WAL and WL, which are caused by quantum interference correction to the resistance. At low field, Fig 3(d) shows a decreasing resistance as the samples is cooled down, indicating that WL is not dominant. The 2D Hikami-Larkin-Nagaoka (HLN) model is commonly used to explain WAL and WL, and is given by[35,36]

$$\Delta G(B) = -\frac{Ne^2}{2\pi h}\left[\psi\left(\frac{B_\varphi}{B}+\frac{1}{2}\right)-\ln\left(\frac{B_\varphi}{B}\right)\right] - \frac{Ne^2}{\pi h}\left[\psi\left(\frac{B_{SO}+B_e}{B}+\frac{1}{2}\right)-\ln\left(\frac{B_{SO}+B_e}{B}\right)\right]$$

$$+ \frac{3Ne^2}{2\pi h}\left[\psi\left(\frac{(\frac{4}{3})B_{SO}+B_\varphi}{B}+\frac{1}{2}\right)-\ln\left(\frac{(\frac{4}{3})B_{SO}+B_\varphi}{B}\right)\right] \quad (1)$$

Where $\psi$ is the digamma function, $B_i = \frac{\hbar}{4eL_i^2}$, $L_\varphi$ is the phase coherence length; $L_{SO}$ is the spin-orbit relaxation length; and $L_e$ is the elastic scattering length. N represents the total number of channels, usually representing subband or valley degeneracy. For samples with low mobility and short elastic scattering length, $B_e \gg B_{SO}, B_\varphi$, and the middle term containing $B_e$ in Eq. (1) is negligible. In TCIs and PbTe, four-fold valley degeneracy can enhance the (anti)localization effect by increasing channel numbers (N≈4). [37,38]

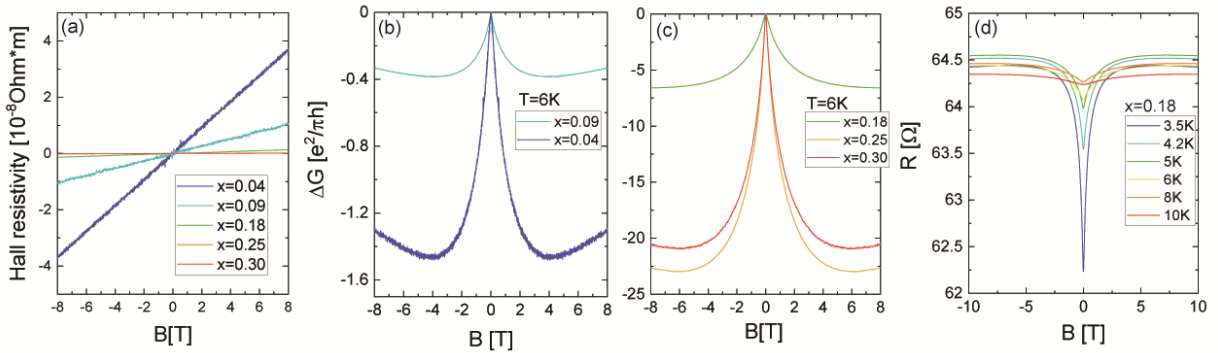

Figure 3. (a) Hall resistivity for samples with different In concentration. As the In content increase the slope of the line gradually decreases. (b,c) Magneto-conductance of samples with different In concentration at 6K, in the normal state. (d) Resistance of sample x=0.18 versus magnetic field at different temperatures.

However, the absolute value of the negative magnetoconductance cusp is high and can reach $10e^2/h$, as shown in figure 3(c). There is a 20 fold–enhancement of its magnitude for x>0.18 which cannot be simply attributed to the WAL effect that typically yield corrections on the order of $1e^2/h$.[35,38,39,40,41] This is especially the case at low temperature below 6K (T/T$_c$<2) as can be seen for x=0.18 in Fig. 3(d). Therefore, superconducting fluctuations must be contributing to the conductance above T$_c$. We consider three most common terms to describe fluctuations. (i) The Aslamazov-Larkin(AL) fluctuations are directly connected to the fluctuating cooper pairs and dominates at T very close to T$_c$ (T/T$_c$<<1).[42,43] (ii) The Maki-Thompson (MT) fluctuations originate from the interaction of electrons with fluctuating Cooper pairs.[44–46] (iii) The density-of-states(DOS) term arises from the change of single-particle density of states due to their involvement in fluctuation pairing.[47,48] As the magnetic field increases, the fluctuations are suppressed, leading to negative magneto-conductivity from AL and MT and a positive contribution from DOS. For 2D superconductors the conductance correction from the MT term is given by:

$$\Delta G_{MT}(T,H) = -\frac{e^2}{\pi h}\beta_L\left(\frac{T}{T_c}\right)\left[\Psi\left(\frac{1}{2}+\frac{B_\varphi}{B}\right) - \ln\left(\frac{B_\varphi}{B}\right)\right] \quad (2)$$

$\beta_L(\frac{T}{T_c})$ is Larkin's electron-electron interaction strength parameter tabulated in Ref. [44]. The MT term has the same form as weak anti-localization in the strong spin-orbit limit but differs by a pre-factor. The AL and DOS terms can be directly calculated at different temperatures, but their contribution is negligible compared to MT at the temperatures of interest.

Our fitting formula is obtained by combining all possible contributions from fluctuations with WL:[49–51]

$$\Delta G(B,T) = -\frac{\left(N+\beta_L(\frac{T}{T_C})\right)e^2}{2\pi h}\left[\psi\left(\frac{B_\varphi}{B}+\frac{1}{2}\right) - \ln\left(\frac{B_\varphi}{B}\right)\right] + \frac{3Ne^2}{2\pi h}\left[\psi\left(\frac{4B_{SO}/3+B_\varphi}{B}+\frac{1}{2}\right) - \ln\left(\frac{4B_{SO}/3+B_\varphi}{B}\right)\right]$$
$$+ \Delta G_{AL}(B,T) + \Delta G_{DOS}(B,T) \quad (3)$$

Notice that Eq. (3) only has $N, B_\varphi$ and $B_{SO}$ as fitting parameters. $B_\varphi$ is temperature dependent and $B_{SO}$ should be temperature independent. $\Delta G_{AL}$ and $\Delta G_{DOS}$ do not contain any fit parameters and are shown to have a very small contribution (see supplement S3) compared to MT. By fitting magneto-conductance at high temperature, where the WL term dominates, we can get a reliable value of $B_{SO}$ and use it to fit the low temperature data. Then, we only have to carry out a two-parameter fitting as a function of temperature. Curve fits for three characteristic samples (x=0.04, 0.18 and 0.3) are shown in figure 4(a)-(c). Weak localization, antilocalization and the Maki-Thompson fluctuations dominate in the temperature range of interest.

Note that the Larkin expression for the MT term is limited to temperatures much higher than T$_c$ and not-too-large fields ($2\pi k_B(T-T_C) \gg \hbar/\tau_\varphi$ and $2\pi k_B(T-T_C) \gg 4eDH$). This limits our fitting range of the data, especially at low temperature close to $T_c$. The fit parameters $N$ and $L_\varphi$ are plotted versus temperature in Fig. 4(d, e) and $L_{SO}$ is plotted versus In content $x$ in Fig. 4(e). Their variation is discussed next.

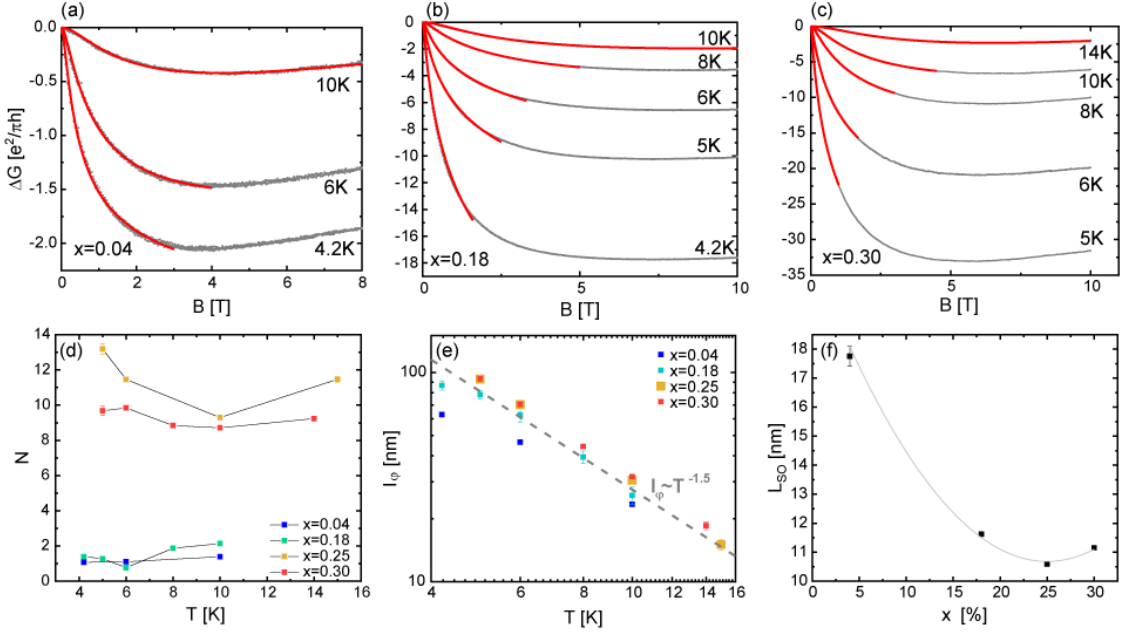

Figure 4. Fitting results for superconducting samples. (a)-(c) show the fitting curves using Eq. (3) for x=0.04, x=0.18 and x=0.30 respectively. (d) Channel number and (e) coherence length versus temperature for samples with different In concentration. (f) Spin-orbit scattering length versus In content x.

The channel number N (Fig. 4(d)) does not exceed 3 for x=0.04 and x=0.18. This is expected in IV-VI materials due to their four-fold valley degeneracy, regardless of topological character.[37,52] Despite taking into account the contribution of fluctuations, N greatly increases as the In concentration goes up. The effect of In on the energy band structure is still unclear in $Sn_{1-x}In_xTe$. Ref.[23] has suggested that In 5s states form an impurity band around Fermi level and generate quasi-localized electrons, which gradually delocalize with increasing x. This idea was supported by transport and NMR measurement on $Sn_{1-x}In_xTe$ crystals[34,53]. In could impact the valley degeneracy by altering the shape of the Fermi surface of SnTe, however, the large N could be a result of our analysis underestimating the magnitude of superconducting fluctuations, especially in samples with $T_c$>3K. The coherence length for all samples roughly follows $L_\varphi \sim T^{-1.5}$. Theoretically, in 2D, inelastic electron-electron scattering will lead to $L_\varphi \sim T^{-0.5}$,[54,55] while electron-phonon scattering yields $L_\varphi \sim T^{-1.5}$.[56,57,58,35] The decay of $L_\varphi$ versus temperature in Fig. 4(e), indicates that electron-phonon scattering is the dominant dephasing mechanism.

We further corroborate the interpretation of the quantum coherent corrections by carrying a similar analysis to the resistance versus temperature R(T) curve of x=0.3 near $T_c$. The detailed expression of weak localization and terms describing superconducting fluctuations are shown in the supplement. The R(T) correction due other fluctuation terms can be calculated from parameters already determined previously (see Supplement S3).[35,41,47,59,60,61] All yield a minimal impact at the temperatures of interest except the MT term. Fig. 5(a) shows a comparison between the R(T) data for x=0.3 and the combined effect of the MT fluctuations and weak localization, qualitatively matching the data (more details are given in the supplement). The MT fluctuations that lead to Eq. (3) are however not expected to appear in measurements of the magnetization as they are related to coherent electron scattering corrections that only influence the conductivity.[62,63] Fig. 5(b) shows AC susceptibility $\chi$ of $Sn_{1-x}In_xTe$ with x=0.3 up to a temperature of 10K at zero magnetic field. There is no visible deviation observed from background diamagnetism of the substrate down to below T=5K, despite R(T) exhibiting a decrease of more than 1% between 10K and 5K. This is consistent with the Maki-Thompson fluctuations model. This observation also rules out the formation of superconducting clusters above $T_c$ as they would normally produce a drop in both R(T) and $\chi(T)$.

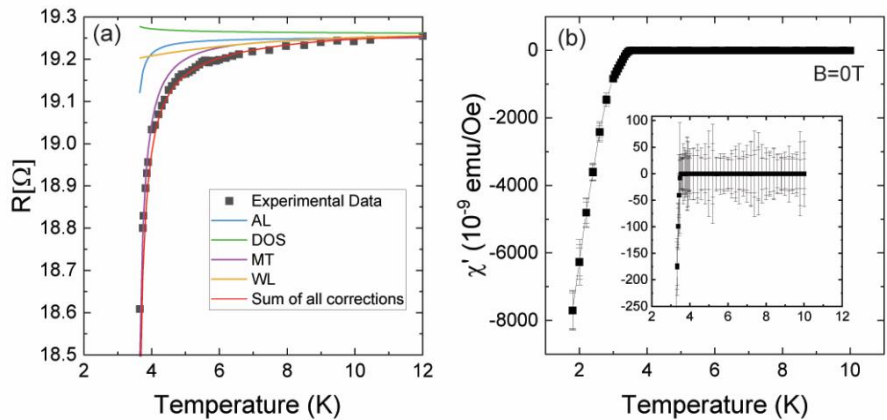

Figure 5. (a) Plot of resistance versus temperature and comparison with different the AL, MR and DOS and WL contributions near $T_c$. Almost all parameters are taken from magneto-conductance fitting result. Detailed expression can be found in the supplement. (b) AC susceptibility vs temperature for *x=0.3*.

Our results thus indicate that $Sn_{1-x}In_xTe$ thin films host strong quantum coherent corrections to the conductivity, originating from the coexistence of superconducting fluctuations and partial antilocalization. We note that a strong WAL has been reported in superconducting $Sn_{1-x}In_xTe$ platelets,[64] but the corrections from superconducting fluctuations were not considered in the literature to the best of our knowledge. Properties of these fluctuations in the normal state may contain crucial information of the superconducting order parameter, as was widely studied in the in cuprate superconductors.[63] Our observation and analysis of the fluctuations thus lay important groundwork for future theoretical and experimental studies on the pairing symmetry of strained $Sn_{1-x}In_xTe$. For the WAL contribution, we attribute it to bulk electrons with spin-orbit splitting instead of the Dirac surface states. This is because on one hand, the WAL cusp is a classical sign of strong spin-orbit interactions that reduce the backscattering probability. On the other hand, we expect the spin-momentum locking of the Dirac surface states to suppress all backscattering, which eliminates WL entirely.[65] However, we find that WAL and WL coexist in our superconducting samples, indicating that carriers from trivial states, that can be spin-orbit split, play a dominant role in normal state transport, and possibly in the superconducting state.

This spin-orbit splitting could originate from several possible sources. For instance, in a recent work on $Pb_{1-x}Sn_xSe$ quantum well, the origin of spin-orbit coupling and weak antilocalization is attributed to surface inversion symmetry breaking that introduces a Rashba-like term in the Hamiltonian describing the system.[66] Band bending [67] can also lead to spin-orbit splitting and can be introduced in a controlled way by growing quantum wells with asymmetric barriers.[68] A ferroelectric crystalline distortion [69,70] that displaces the Sn and Te sublattice with respect to each other is known to occur in SnTe and can also yield a strong three-dimensional Rashba splitting by breaking bulk inversion symmetry[71,72]. We have carried out a systematic search for this distorted state using low temperature X-ray diffraction measurements and resistivity measurements, but did not find any evidence of it in $Sn_{1-x}In_xTe$ (see supplement S4).[73,74,75,76] Structurally, the (513) RSM in Fig. 1(e) shows an in-plane compressive strain and out-of-plane tensile strain. This can break the cubic symmetry yielding a rhombohedral distortion. But, it preserves three-fold rotational symmetry in the film plane and centroymmetry since it is not accompanied by a sublattice shift. While several routes can cause a Rashba splitting in IV-VI materials, to identify the precise origin of the spin-orbit splitting in our samples requires further experiments.

In strongly spin-orbit-coupled systems, such as $Cu_xBi_2Se_3$, $Sn_{1-x}In_xTe$, and various transition metal dichalcogenides, parity-mixed superconductivity can naturally exist on symmetry grounds.[77,78,79] Odd-parity or sometimes topological superconductivity can also be energetically favored by spin-orbit coupling even when mediated by phonons[8,79,80,77]. We thus expect $Sn_{1-x}In_xTe$ thin films to be a plausible platform hosting unconventional superconductivity.

In summary, we have grown $Sn_xIn_{1-x}Te$ (0<x<0.3) thin films by MBE and measured their structural, quantum coherent and superconducting transport properties. Samples can maintain superconductivity up to 3.5K with x=0.3. In the normal state, we have observed the coexistence of quantum coherent corrections from WAL, WL and superconducting fluctuations. Their analysis of the coherence length indicates yields a short spin-orbit scattering length consistent with strong spin-orbit splitting. The latter is a necessary but not sufficient ingredient to realize topological superconductivity. To the best of our knowledge, superconducting fluctuations have not been observed in a topological insulator that goes superconducting. Our results thus raise several interesting questions about the superconductivity of $Sn_xIn_{1-x}Te$ thin films. First, they motivate further theoretical studies of $Sn_{1-x}In_xTe$ with the distorted crystal structure to elucidate the nature of its superconductivity. Second, they raise the question of whether superconducting fluctuations and their response to magnetic field and impurities can shed light on the pairing symmetry. At the very least, the synthesis of such thin films provides a superconducting material that one can use to grow in-situ proximity effect bilayers with epitaxial interfaces based on TCIs.

FUNDING SOURCES

Work supported by NSF-DMR-1905277. LPR acknowledges support from NSF-DMR-2005092 The Material Characterization Facility is funded by the Sustainable Energy Initiative (SEI), which is part of the Center for Sustainable Energy at Notre Dame (ND Energy). We also acknowledge support from the Notre Dame Integrated Imaging Facility. This research used resources of the Advanced Photon Source, a U.S. Department of Energy (DOE) Office of Science User Facility, operated for the DOE Office of Science by Argonne National Laboratory under Contract No. DE-AC02-06CH11357.

ACKNOWLEDGMENT

B. Janko and M.R. Eskildsen for useful discussions about superconducting fluctuations. We thank Neil R. Dilley for carrying out the magnetic susceptibility measurements.